%% file: sub-1768.tex
\documentclass[sigconf]{acmart}
\AtBeginDocument{%
  \providecommand\BibTeX{{%
    \normalfont B\kern-0.5em{\scshape i\kern-0.25em b}\kern-0.8em\TeX}}}

\copyrightyear{2024}
\acmYear{2024}
\setcopyright{rightsretained}
\acmConference[DIS '24]{Designing Interactive Systems Conference}{July 1--5, 2024}{IT University of Copenhagen, Denmark}
\acmBooktitle{Designing Interactive Systems Conference (DIS '24), July 1--5, 2024, IT University of Copenhagen, Denmark}\acmDOI{10.1145/3643834.3661577}
\acmISBN{979-8-4007-0583-0/24/07}

\usepackage{subcaption, graphicx, enumitem}

\begin{document}


\title[\wbname{}]{\wbname{}: A Workbook to Support Effective Collaboration Between AI Experts and Clients}

\author{Dae Hyun Kim}
\email{dhkim16@cs.stanford.edu}
\orcid{0000-0002-8657-9986}
\affiliation{%
  \institution{KAIST}
  \streetaddress{291 Daehak-ro, Yuseong-gu}
  \city{Daejeon}
  \country{Korea}
  \postcode{34141}
}

\author{Hyungyu Shin}
\email{hyungyu.sh@kaist.ac.kr}
\orcid{0000-0001-7328-0072}
\affiliation{%
  \institution{KAIST}
  \streetaddress{291 Daehak-ro, Yuseong-gu}
  \city{Daejeon}
  \country{Korea}
  \postcode{34141}
}

\author{Shakhnozakhon Yadgarova}
\email{yadgarova@kaist.ac.kr}
\orcid{0000-0003-3299-8867}
\affiliation{%
  \institution{KAIST}
  \streetaddress{291 Daehak-ro, Yuseong-gu}
  \city{Daejeon}
  \country{Korea}
  \postcode{34141}
}

\author{Jinho Son}
\email{sjhfam@algorithmlabs.co.kr}
\orcid{0000-0001-7403-0390}
\affiliation{%
  \institution{Algorithm Labs}
  \streetaddress{18 Maebongsan-ro, Mapo-gu}
  \city{Seoul}
  \country{Korea}
  \postcode{03911}
}

\author{Hariharan Subramonyam}
\email{harihars@stanford.edu}
\orcid{0000-0002-3450-0447}
\affiliation{%
  \institution{Stanford University}
  \streetaddress{450 Jane Stanford Way}
  \city{Stanford}
  \state{CA}
  \country{USA}
  \postcode{94305}
}

\author{Juho Kim}
\email{juhokim@kaist.ac.kr}
\orcid{0000-0001-6348-4127}
\affiliation{%
  \institution{KAIST}
  \streetaddress{291 Daehak-ro, Yuseong-gu}
  \city{Daejeon}
  \country{Korea}
  \postcode{34141}
}

\newcommand{\wbname}[0]{\textsc{AINeedsPlanner}}

\renewcommand{\shortauthors}{Kim et al.}

\newcommand{\change}[1]{{{#1}\normalfont}}
\newcommand{\cut}[1]{{\color{lightgray}{#1}\normalfont}}

\input{sections/00-abstract}

\ccsdesc[500]{Software and its engineering~Collaboration in software development}
\ccsdesc[300]{Computing methodologies~Artificial intelligence}

\keywords{AI application planning, client-AI expert collaboration, information flow, workbook}


\received{20 February 2007}
\received[revised]{12 March 2009}
\received[accepted]{5 June 2009}

\maketitle

\input{sections/01-intro}
\input{sections/02-related-work}
\input{sections/03-interview}
\input{sections/04-gap-study}
\input{sections/05-workbook}
\input{sections/06-discussion}
\input{sections/07-limitations-future-work}
\input{sections/08-conclusion}

\begin{acks}
\change{
The authors thank Yoonsu Kim and Nicole Lee for their early contributions during the formative interview and initial explorations of the workbook format.
The authors also thank the AI engineers at AlgorithmLabs and the members of KAIST Interaction Lab for their constructive feedback throughout the work and piloting the formative interview and the main study.
This work is supported by the G-CORE Research Project grant at KAIST.
Hariharan Subramonyam is supported by NSF award IIS-2302701.}
\end{acks}

\bibliographystyle{ACM-Reference-Format}
\bibliography{bibliography/methodology,bibliography/tools,bibliography/ml-methods,bibliography/intro,bibliography/discussion,bibliography/related-work,bibliography/future-work,bibliography/additional,bibliography/limitations}


\end{document}

%% file: sections/00-abstract.tex
\begin{abstract}
Clients often partner with AI experts to develop AI applications tailored to their needs. In these partnerships, careful planning and clear communication are critical, as inaccurate or incomplete specifications can result in misaligned model characteristics, expensive reworks, and potential friction between collaborators. Unfortunately, given the complexity of requirements ranging from functionality, data, and governance, effective guidelines for collaborative specification of requirements in client-AI expert collaborations are missing. In this work, we introduce \wbname{}, a workbook that AI experts and clients can use to facilitate effective interchange of clear specifications. The workbook is based on (1) an interview of 10 completed AI application project teams, which identifies and characterizes steps in AI application planning and (2) a study with 12 AI experts, which defines a taxonomy of AI experts' information needs and dimensions that affect the information needs. Finally, we demonstrate the workbook's utility with two case studies in real-world settings.
\end{abstract}

%% file: sections/01-intro.tex
\section{Introduction}
Creating AI applications that are human-centered, ethical, and meet regulatory requirements requires effective communication and collaboration between different stakeholders with diverse backgrounds, incentives, and values~\cite{delgado2021stakeholder,van2018prototyping}.
Because of the complexity of the AI technology~\cite{davenport2018artificial,gartner2019gartner,nam2021adoption,yang2020re}, we have seen the separation of those who desire to introduce AI into their workflow and AI experts who have the ability to build AI models.
Hence, \textit{client-AI expert collaborations} has become a widespread form of collaboration between stakeholders~\cite{statistica2023artificial}.

Client-AI expert collaborations, which can either take \textit{inter-organization} forms or \textit{inter-team} forms, are characterized by the large amounts of discussion during the planning stage prior to execution by the AI experts~\cite{piorkowski2021how, subramonyam2021can}.
During this discussion, the clients should not only determine and share their goals, expectations, and available resources, but also share additional information, including domain knowledge around the data~\cite{veale2018fairness, vertesi2011value} and domain-specific performance metrics~\cite{holstein2019improving, shi2020artificial}, for a successful AI application development.
\change{Through a formative interview of 10 successfully completed AI application development projects (Section~\ref{sec:interview}), we further find that the planning stage of AI application development is innately an information flow from the client to the AI expert through three phases; (1) the client initially define their goals and assess their needs and resources before entering collaboration, (2) the client passes on the information with the AI expert while undergoing discussions to iterate on the information, and (3) the AI expert makes execution plans based on the iterated information.}

Despite the need to closely involve the clients, many client-AI expert collaborations encounter significant challenges.
To begin with, traditional engineering methods have not been designed for sufficiently involving the clients, despite the importance of information flow from the client~\cite{sloane2022participation, tan2023seat}.
To make things worse, the knowledge barriers between the client and the AI expert only add to the challenges~\cite{bogina2021educating, estivill2022constructing, williams1993translation, subramonyam2022solving}; clients not only have difficulty forming a mental model of the information needs for development to start, but also expressing information in forms that can be readily incorporated into the AI application design by the AI experts.
\change{The need to closely involve clients and the knowledge barriers between the clients and AI experts exacerbate the challenges in determining the uniquely important aspects of AI application development, such as datasets and success measures~\cite{google2019people, mitchell2019model, tan2023seat}.
For instance, consider the planning process for an AI-powered cancer treatment recommendation tool for doctors. 
Despite the dataset's importance, doctors may struggle to figure out the expected data format and quantity for training the AI model. 
Similarly, AI experts may face challenges in determining available data and relevant fields. 
Moreover, doctors may have vague goals, making it difficult to express them in quantifiable metrics for AI development.
Discussions on these topics will require overcoming the knowledge barrier that entails substantial effort from both stakeholders due to the high degrees of specialization of the two fields.
}

\begin{figure}
    \centering
    \includegraphics[width=\linewidth]{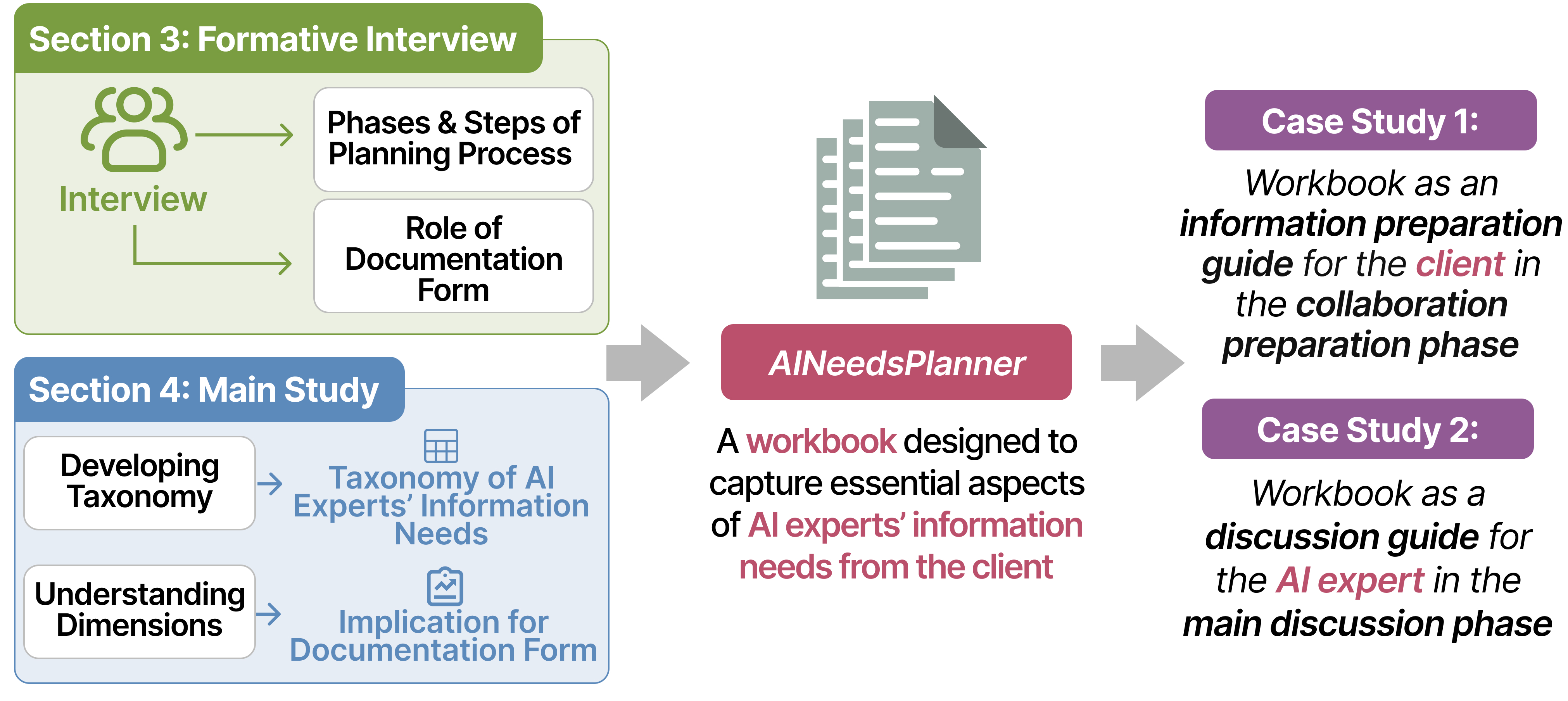}
    \caption{An overview of the structure of the paper. We use the \change{formative} interview and the \change{main} study results to inform the \wbname{} workbook, on which we performed two case studies.}
    \Description{Diagram illustrating the structure of the paper. On the left, two sections are outlined: 'Section 3: Formative Interview', colored green, which includes sub-points 'Phases and Steps of Planning Process' and 'Role of Documentation Form'; directly below, 'Section 4: Study', colored blue, which includes 'Developing Taxonomy' and 'Understanding Dimensions'. Grey-colored arrows connect these sections to a central document icon representing the 'AINeedsPlanner', a workbook designed to capture essential aspects of AI experts' information needs from the client. A second arrow extends to the right, leading to two case studies presented in a purple color scheme. 'Case Study 1' is positioned at the top right, highlighting the workbook as an information preparation guide for the client during the collaboration preparation phase, while 'Case Study 2' focuses on the workbook as a discussion guide for the AI expert in the main discussion phase.}
    \label{fig:overview}
\end{figure}

Based on these existing challenges \change{and the widespread adoption of documentation during the planning stage~\cite{piorkowski2021how}}, we introduce the \wbname{} workbook as a tool for facilitating effective collaboration between AI experts and their clients with deeper client involvement (Figure~\ref{fig:overview}).
\change{Based on the needs we uncovered during the formative interview, we design the workbook as a versatile form that can not only be used as a guide for the client as they define their goals and assess their needs and resources prior to discussions but also as a discussion roadmap during the client-AI expert discussion.
To define the contents of this workbook, we conducted a study with 12 AI experts to taxonomize and characterize the information needs of the AI experts from their clients.
To understand the value of \wbname{} as an information preparation guide for the client while preparing for collaboration and a discussion guide and checklist, we put the workbook to use in two different case studies.
}

The contributions of this paper are as follows:
\begin{itemize}
\item Identification and characterization of constituent steps of the AI application planning in client-AI expert collaboration;
\item A taxonomy of AI experts' information needs from clients during AI application planning;
\item \wbname{}, a workbook for supporting client-AI collaboration in the planning stage; and
\item Two case studies providing insights into the value of \wbname{} as both an information preparation guide for the client and a discussion guide and checklist.
\end{itemize}

%% file: sections/02-related-work.tex
\section{Related Work}
Our work is grounded in three main areas of prior work: (1) Supporting AI specification, (2) Workflows for building AI applications, and (3) Collaboration for building AI applications.

\subsection{Supporting AI Specification}

Building AI applications is a challenging task that involves a large number of decisions with multiple expertise needed. Since it requires a significant amount of time and effort to build, the process needs careful planning that clearly defines the goal and identifies necessary resources to make sure that the resulting AI application creates value. However, little research has investigated designing specifications for AI application planning. Existing specification templates such as Datasheets for Dataset~\cite{gebru2021datasheets} and Model cards~\cite{mitchell2019model} can be used in the planning stage for clearly setting the design goals of AI applications, but the nature of ``reporting'' as the goal misses important dimensions that need to be considered in the planning stage (e.g., available resources for training models and expected impact of the application to the organization). More importantly, the templates assume some level of AI expertise by using AI-related terminology, which makes it challenging for clients (i.e., AI novices) to fill in the information in the planning stage.

To support lay users in designing AI applications, research introduced guidelines, more hands-on workbooks, and interactive systems. The guidelines~\cite{google2019people, amershi2019guidelines, microsoft2022responsible, apple2022human, IBM2021infuse, subramonyam2022solving, heer2019agency, wang2020human, Humancen86:online} inform a set of important considerations with specific examples. When it comes to clients planning AI applications, however, it is challenging to contextualize the guidelines for a specific context because of a lack of AI experiences. People+AI Guidebook~\cite{google2019people} provides fill-in-the-blank exercises that scaffold the process across six chapters, using simple languages. However, it is unclear whether they are complete as the planning material for supporting the early stage~\cite{yildirim2023investigating}. In this work, we aim to (1) identify a set of information needed in the planning stage and (2) create a workbook that elicits the information from the clients so that the clients can be well-prepared for building AI applications.

\subsection{Workflows for Building AI Applications}

The process of building AI applications involves multiple stages where some of the important tasks are (1) setting clear goals about what AI should produce and defining success metrics, (2) designing and collecting datasets, and (3) building and evaluating models. Due to the uncertain and complex nature of AI, research showed that the AI application-building process entails unique challenges across the design process~\cite{yang2020re}. To address such challenges, research introduced interactive systems supporting various stages (e.g., data analysis~\cite{bauerle2022symphony}, prototyping~\cite{subramonyam2021protoai}, and UI building for multiple stakeholders~\cite{franccoise2021marcelle}). However, it is not yet investigated how to systematically support planning AI applications, especially for the clients.

Planning is a crucial step for determining the goals and means of achieving the goal~\cite{nash2013nature}. Prior work in various domains shows that planning can significantly increase the performance of organizations and quality of work~\cite{delmar2003does, schendel1979new, saddler2004preventing, limpo2018effects}. For the AI application building process, research also highlighted the planning phase as a separate step~\cite{subramonyam2021towards} and noted the stage as a long preparatory stage~\cite{yang2018investigating}. However, it remains unclear what the workflow of the planning is as well as the underlying challenges. As clients are knowledgeable in their domain and contexts, they have the potential to make unique contributions in the planning stage such as clearly defining what AI should do, designing domain-specific target metrics, estimating data availability, and depicting interactions between the user and AI to make it usable in real-world contexts. But to make such a contribution, clients need to have concrete guidance that elicits such information from the clients. We aim to design a workbook for eliciting such information from the clients so that the workflow of the planning gets clearer and more systematic.

\subsection{Collaboration for Building AI Applications}

As the process of building AI applications is complicated, it involves a collaborative approach between multiple stakeholders (e.g., domain experts, AI experts, and UI/UX designers). However, it is challenging to support such collaboration largely due to the complexity of AI~\cite{yang2020re}. For example, designers find it challenging to understand the capabilities of AI models and their behaviors~\cite{dove2017ux, subramonyam2022solving}. In making decisions, domain experts require information more than prediction results of AI models~\cite{liao2020questioning, cai2019hello}. To address such challenges, research suggested improved methods for collaboration~\cite{subramonyam2021towards, subramonyam2022solving}. For instance, research proposed a process model for collaboration that utilizes user data as probes~\cite{subramonyam2021towards}.

However, little research has investigated the collaboration between clients and AI experts for planning AI applications. Research showed that communications between different stakeholders entail challenges in establishing common grounds~\cite{mao2019data}. Boundary objects can resolve the challenge~\cite{carlile2002pragmatic}, but such boundary objects for the planning AI application have not been designed yet. Similar to prior research introducing various artifacts as boundary objects (e.g., \change{checklist~\cite{bharadwaj2019critter}, action plan~\cite{rahman2020mixtape}}, data documentation~\cite{heger2022understanding}, AI explanations~\cite{ayobi2021machine}, and UI prototype ~\cite{subramonyam2021protoai}), we design a workbook as a boundary object for the planning stage by understanding information that needs to be shared between clients and AI experts as well as underlying challenges during the planning stage.

%% file: sections/03-interview.tex
\section{Formative Interview: Understanding the Planning Process between AI Experts \& Clients}
\label{sec:interview}
To understand the process of AI application planning between an AI expert and their clients as well as how documentation forms could fit in to further facilitate this process, we performed a semi-structured interview of clients and AI experts who have participated in successfully completed AI application development projects.
In the \change{formative} interview, we specifically sought to find answers to the following research questions:

\vspace{0.025in}
\noindent[\textbf{RQ1}] What steps constitute the AI application planning process between an AI expert and their client and what are the key characteristics of these steps?

\vspace{0.025in}
\noindent[\textbf{RQ2}] How can a form of documentation fit into the AI application planning process in client-AI expert collaboration to further facilitate the process?

\subsection{Participants}
\begin{table}[t]
    \caption{An overview of AI application projects we surveyed.}
    \Description{Table providing an overview of AI application projects surveyed, categorized by project identifier, project structure, project domain, and participant role. Rows are divided into two main sections: 'Inter-Organization' and 'Inter-Team'. The 'Inter-Organization' projects include IO1 in Education, IO2 in Food and Manufacturing, IO3 in Pharmaceutical, and IO4 in Robotics and Cooking, with varying roles of Client and AI Expert. The 'Inter-Team' projects, listed as IT1 through IT6, span domains such as Education, Commerce and Video, Medical, Language, Finance and Banking, and a combination of Medical and Pharmaceutical, with roles assigned to either Client or AI Experts. }
    \footnotesize{
    \begin{tabular}{|l|l|l|l|}
    \hline
    Project & Project Structure & Project Domain & Participant Role \\ \hline \hline
    IO1 & Inter-Organization & Education & Client, AI Expert \\
    IO2 & Inter-Organization & Food, Manufacturing & Client, AI Expert \\
    IO3 & Inter-Organization & Pharmaceutical & Client, AI Expert \\
    IO4 & Inter-Organization & Robotics, Cooking & Client \\
    IT1 & Inter-Team & Education & Client \\
    IT2 & Inter-Team & Commerce, Video & Client \\
    IT3 & Inter-Team & Medical & AI Expert \\
    IT4 & Inter-Team & Language & AI Expert \\
    IT5 & Inter-Team & Finance, Banking & AI Expert \\
    IT6 & Inter-Team & Medical, Pharmaceutical & AI Expert \\
    \hline
    \end{tabular}
    }
    \label{tab:organization}
\end{table}

To obtain detailed perspectives from the two stakeholders of interest, we recruited (1) \textit{AI experts} and (2) \textit{clients} of recent AI application projects that have concluded successfully (self-reported), a restriction put to capture the good planning practices.
We required that the participant took part in the planning phase of the AI application project.

We focused our interviews on \textit{inter-organizational projects} (IO1-4) as well as \textit{inter-team projects} within the same organization (IT1-IT6).
We excluded inner-team projects or projects motivated by the AI experts themselves as they display no AI expert-client relationship.
\change{The AI experts were typically AI engineers, researchers, or technical leads of the AI companies (e.g., CTO, team leader), while the clients ranged from employees in client teams or companies to management roles at client teams or companies (e.g., project manager/owner, CEO).}

We recruited the participants through the authors' personal and academic connections and then employed snowball sampling~\cite{goodman1961snowball} via referrals.
While recruiting, we focused on capturing diverse project domains by limiting ourselves from including more than two projects from the same domain.
In addition, to respect any secrecy regulations of the participants' organizations, we informed the participants about the contents of the interview beforehand.
While we recorded the interviews by default, we also offered the option of not recording if it goes against any of the policies of the participant's organization.

\subsection{Procedure}
We performed the interviews on Zoom~\cite{zoom2022online} to abide by various COVID-related regulations and to allow ourselves to reach out to participants outside of our immediate geographical vicinity.

At the beginning of the interview, we introduced ourselves and provided an overview of the contents of the interview.
We then asked for general information about the participant's organization and the domain of the AI application development project we would discuss during the interview.
Next, we conducted the main interview, where we asked about the procedures and the contents of planning.
While asking them to describe the process of planning, we asked about the stakeholders involved in the process and their roles, the information flow between the stakeholders, and various characteristics of the process including the challenges and the remedies.
Finally, we allowed the participants to freely share their thoughts and ask us any questions they had throughout the interview.

For analysis, we first extracted each step in the process mentioned for each of the projects in Miro~\cite{miro2024}.
We then used an affinity diagram~\cite{kawakita1982original,kawakita1986kj} on these steps to come up with steps that commonly appear throughout multiple projects.
After identifying the common steps, we referred back to the original interviews to further group these steps into phases and to understand the relative ordering of these steps.

\subsection{Assessing RQ1: The Process of Planning}

\begin{figure}[t]
    \centering
    \includegraphics[width=\linewidth]{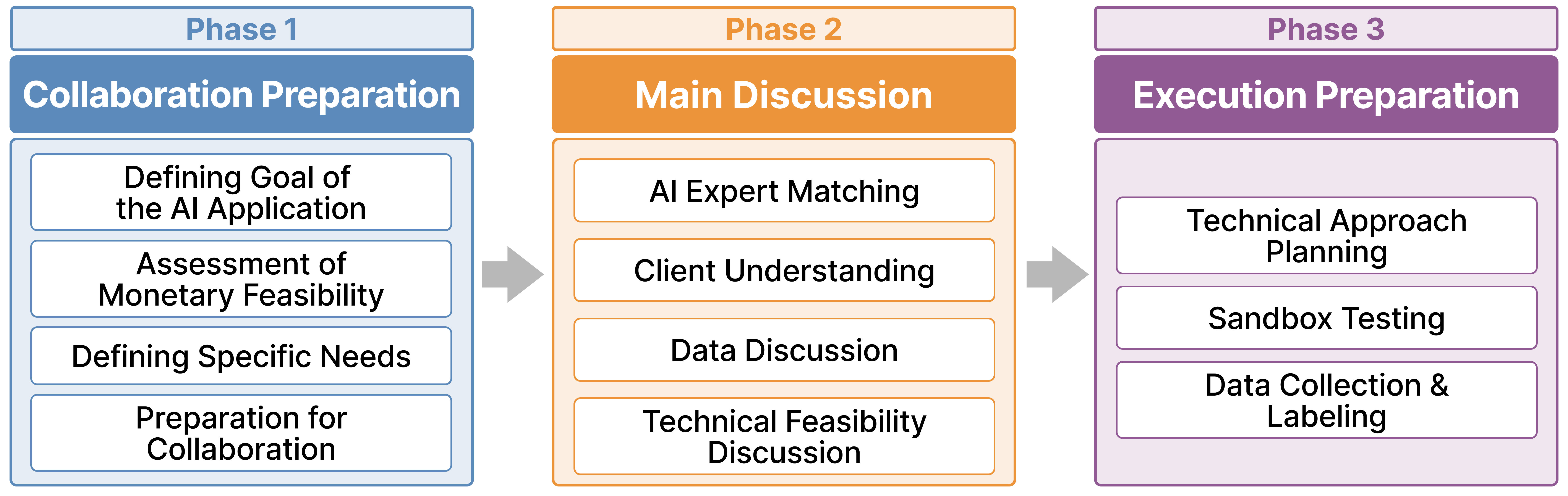}
    \caption{The three phases of the AI application planning process in client-AI expert collaboration and the comprising steps of each of the phases.}
    \Description{This figure represents a three-part horizontal flowchart that outlines the phases of AI project planning. 'Phase 1: Collaboration Preparation', colored blue and located to the left, includes tasks such as 'Defining the Goal of the AI Application', 'Assessment of Monetary Feasibility', 'Defining Specific Needs', and 'Preparing for Collaboration'. A gray arrow then points to the next section, 'Phase 2: Main Discussion', colored orange, which includes 'AI Expert Matching', 'Client Understanding', 'Data Discussion', and 'Technical Feasibility Discussion'. Another gray arrow leads to the final phase, 'Phase 3: Execution Preparation', colored purple, which involves 'Technical Approach Planning', 'Sandbox Testing', and 'Data Collection and Labeling'. The phases are interconnected with gray arrows, indicating the sequential progression from preparation through discussion to execution.}
    \label{fig:planning-process}
\end{figure}
Through our analysis, we identified a total of 11 steps within the planning and grouped them into 3 phases: (1) the collaboration preparation phase, (2) the main discussion phase, and (3) the execution preparation phase  (Figure~\ref{fig:planning-process}).

\subsubsection{Phase 1. Collaboration Preparation Phase}
Before the collaboration with an AI expert begins, \change{planning begins on the client-side. A} client typically (1) defines the goal they wish to achieve with the AI application, (2) confirms whether the AI application would make 'financial sense', (3) defines any specific needs for the model, and (4) performs initial preparations before entering a collaboration with AI expert.

\vspace{0.05in}
\noindent\textit{\textbf{Defining Goal of the AI Application.}}
One of the most important tasks that the client must perform during this phase is to define the goal of the AI application.
In most if not all cases, the goal includes the problem that the client wishes to solve with the application, the target users of the application, and how the application is to be used, although the level of specificity may differ.
Although not common, the client may occasionally possess concrete target measures for success (IT1).

\vspace{0.05in}
\noindent\textit{\textbf{Assessment of Financial Feasibility.}}
Before seeking AI teams and kick-starting the AI application development, the client needs to confirm what budget they have available for data collection, development, and deployment and whether introducing the AI application would pay off, especially if the client is part of a for-profit organization (IO2-4, IT1-6).
While the budget for development and deployment is usually the person in managerial roles, the considerations that go into determining profitability vary.
For smaller organizations (e.g., startups), the profitability assessment is often based on the likelihood of securing more investment by distinguishing themselves from their competitors (IT2, 3), government funding for startups applying AI technology in niche domains (IO3, IT2, 6), or a more direct suggestion by the investors to introduce AI into their workflow (IO4).

\vspace{0.05in}
\noindent\textit{\textbf{Defining Specific Needs.}}
Some of the clients have very specific needs about various components of AI application development.
The most common needs that clients have are a timeline or a deadline by which they need the application developed by (IO1, IT1, 6) or specific features to be included in the resulting AI application (e.g., dashboard visualization for IO2, personalization for IO4, explainability for IT5).
There are occasionally more specific needs about the core AI model that would power the AI application, especially an update to a novel AI model (e.g., a specific deep learning-based natural language processing framework for IT3, large language models for IT4, deep learning for IT6).

\vspace{0.05in}
\noindent\textit{\textbf{Preparation for Collaboration.}}
Before starting collaboration with AI experts, some clients perform additional preparation beforehand to streamline the communication and to be able to have more concrete discussions.
Clients occasionally precollect data and preprocess the data into a form that they can share with AI experts in accordance with corporate and regional restrictions (IO1, 3).
Compliance with corporate restrictions in terms of data sharing is more of an issue of inter-organization collaborations than of inter-team collaborations.
Other than data-related preparations, one of the clients went as far as building an evaluation framework before starting collaborations with AI experts (IT1).

\vspace{0.05in}
\noindent\textit{\textbf{Notes on Phase 1.}}
\change{While it would be ideal for the client to have been able to clearly define their goals and needs prior to starting collaboration with AI experts, it is usually unrealistic.}
In particular, clients' lack of AI knowledge and the uncertain nature of AI uncertainties at the time of planning usually limit forming concrete and correct expectations about the AI model that would power the application (IO1), the characteristics of the errors the underlying AI model may give (IT1), and the data quantity and format required to train the underlying AI model (IO4, IT2).
Hence, while the first phase serves an important purpose in specifying the client's initial goal and expectations, these initial goals and expectations are concretized and calibrated through collaboration with AI experts in later phases.

\subsubsection{Phase 2. Main Discussion Phase}
The second phase is where the client's goals, situational information, and expectations formulated in Phase 1 are passed to AI experts so that the AI experts can have sufficient information based on which they can come up with initial approaches for execution.
A notable characteristic of this phase is that while the general direction of information flow is from the clients to AI experts, significant amounts of discussions occur between clients and AI experts, allowing further concretization of the various components determined by the client in Phase 1.

\vspace{0.05in}
\noindent\textit{\textbf{AI Expert Matching.}}
The first step of the main communication phase involves seeking AI experts who can best fulfill their AI application development needs.
The main criteria for determining the fit of the AI expert in the AI application development is their understanding of and experience with the domain of the desired AI application.

\change{For inter-organization projects, the client selects the AI expert best fit for developing their AI application by going through a bidding process (IO1, 2) or by looking at the AI expert's portfolio to gauge prior experience in the domain (IO3, 4). 
For inter-team projects within the same organization, the client team simply asks the team with AI expertise, which naturally possesses both an understanding and experience in the domain, to develop an AI application for them.}

\vspace{0.05in}
\noindent\textit{\textbf{Client Understanding.}}
While the core information passed from the clients to the AI experts is focused on the goal, available resources, and client expectations, AI experts often need to request additional information from the client about the domain and their workflow to make it across the knowledge barrier.
Doing so allows AI experts to carry on discussions with the client about the data and technical feasibility and sketch out potential technical approaches.
Hence, the clients provide descriptions of their domain and their workflow to the AI experts (IO1, 2, 4, IT6).

However, despite the efforts to reduce the knowledge barriers and provide context, many collaboration efforts still end up experiencing some challenges in terms of communication and need to invest additional time and effort to work around the barrier (IO1-4, IT5, 6).

\vspace{0.05in}
\noindent\textit{\textbf{Data Discussion.}}
Data is among the most important pieces when it comes to AI application development not only because it further informs the input and the output, but also because data quality is crucial in determining the performance of the AI model trained on the data.

If the client already has a collected dataset, they would pass the dataset over to the AI expert and discuss whether the data can be worked with or additional data or preprocessing is required (IO3).
In the cases in which the client has not collected data, AI experts and the client arrive at a plan for dataset collection.
\change{They do so} by matching what the AI experts believe they need based on the given goal and what the client can collect and provide to the AI expert (IO1, IT2).
Furthermore, if the label is missing from the data, the client and the AI experts also discuss data labeling methods (IO3).

The client also shares with the AI expert their domain insights on the features in the data that are likely more relevant than others (IO1, 2, IT5).

\vspace{0.05in}
\noindent\textit{\textbf{Technical Feasibility Discussion.}}
Based on the information provided by the client, the AI expert must gauge the feasibility of the goals set by the client given the available resources and expectations.
It may occasionally be the case that the client's goals and expectations exceed what is possible with the available resources.
In these cases, the AI experts need to calibrate their client's goals and expectations to something feasible with the given resources.

Because the clients usually lack AI expertise to perform technical feasibility assessments, they rely on AI experts for the decision (IO2, IT3, 5, 6)

\subsubsection{Phase 3. Execution Preparations}
The execution preparation phase is characterized as a stage in which the AI experts play the central role, although the client may be involved to some degree.

\vspace{0.05in}
\noindent\textit{\textbf{Technical Approach Planning.}}
Based on the information obtained and discussed with the client in Phase 2, the AI experts sketch out ways to meet the provided goals and the client's expectations, while being subject to the various restrictions set by the available resources.

The technical approach planning often involves sandbox testing along the way, especially if prior attempts to apply the planned AI model on the specific problem domain are sparse (IT1-5).
After defining a potential technical approach, AI experts build small prototypes of the AI model often with parts of the data (e.g., minimum viable product) and verify the prototype to check whether they can achieve the proposed goal of the client and whether it meets the client's expectations.
\change{When the success targets are relatively clear, the AI expert can perform this on their own (IT1, 2, 4); when they are unclear, verification of the sandbox testing results occasionally involves the client (IT3, 5).}

\vspace{0.05in}
\noindent\textit{\textbf{Dataset Collection \& Labeling.}}
Clients or AI experts collect data and label to be used during the execution stage if they have not been collected by this phase (IO1, 3, IT1, 4).

\subsection{Assessing RQ2: Documentation Form in Facilitating the Planning Process}
As previously noticed by Piorski et al.~\cite{piorkowski2021how}, many of the projects we surveyed already employed some documentation during the planning stage (IO1-4, IT1, 2).
Yet, the projects either utilized documentation in the absence of a predefined format (IT1) or documentation with only the high-level sectioning (IO1-4, IT2).
The task of determining the exact contents with which to fill in the documentation follows the discretion of two stakeholders based on their tacit knowledge of the required discussions acquired through years of experience.
A standardized documentation form would help centralize and surface the tacit knowledge that is required through the process.
The centralized documentation form would not only help lower the entry barrier to client-AI expert collaboration, but also help the two stakeholders form expectations when initiating client-AI expert with a stakeholder they have not worked with, which is especially common for inter-organization collaborations (IO2-4).

We observed two major ways that the clients and AI experts were employing documentation in planning.
The first use case of documentation was centered around the client during the collaboration preparation phase as they determine what information to consider (IO1-2), and during the main discussion phase, they initially pass on that information to the AI expert (IO1-2).
The second use case of documentation was centered around the AI expert during the main discussion phase as they take notes of the meetings for future references (IT1, 2) and guide discussions with their clients (IO3), and during the execution preparation phase as they formulate a technical approach and communicate the back to the client asynchronously (IO1-2, 4).

Based on these use cases, we deduce that a documentation form that clearly outlines the information needs of the AI experts as they draft a technical approach can be useful for both use cases leading up to the main discussion phase.
In the first use case centered around the client, a detailed documentation form would help clients understand what expectations they should set and assessments they should make prior and guide them through the process, while also serving as a form that they can send to the AI expert without much additional effort.
In the second use case centered around the AI expert, a detailed form can not only help the AI expert organize the discussion, but can also help them ensure that all key information has been discussed with the client.

%% file: sections/04-gap-study.tex
\section{Main Study: Understanding \& Taxonomizing the Information Needs of AI Experts from Clients}
\label{sec:experiment}
While we have established a need for a documentation form outlining the detailed information needs of AI experts from their clients through the \change{formative} interview in Section~\ref{sec:interview}, the steps identified from the interview only provide some higher-level insights about the contents of the documentation form.
Hence, to arrive at a more detailed list of information needs to inform the contents of the documentation form, we conducted a study on 12 AI experts asking for their information needs on AI application development problems collected from the real world.
\change{We form a taxonomy of the information needs of AI experts from their clients through affinity diagramming~\cite{kawakita1982original,kawakita1986kj}, which would form the basis of the contents of our documentation form in Section~\ref{sec:workbook}.}

We performed the study with two research questions:

\vspace{0.025in}
\noindent[\textbf{RQ1}] What is the taxonomy of the information needs from the AI expert to the client during the planning stage?

\vspace{0.025in}
\noindent[\textbf{RQ2}] What affects these information needs?

The study protocol we used is included in the supplemental material.

\subsection{Participants}
\begin{table*}
\caption{An overview of the AI experts who participated in our study and the \change{AI application ideas} they selected for the study \change{with the respective application domains in brackets}.}
\Description{Table summarizing the profiles of AI experts who participated in a study and the selected AI application ideas they worked on, categorized by participant number, years of experience, and a detailed description of the AI application idea. Each row represents a participant, labeled from P1 to P12, with their experience ranging from 1 to 10 years. The AI application ideas are diverse, covering sectors such as Manufacturing, Finance, Education, Art, Productivity, and Military, with specific projects like predicting demand for food products, classifying loan applications, predicting university student dropout, generating cartoon sketches, providing personalized course recommendations, detecting defective pixels in televisions, building automated phone answering systems, manipulating image light sources, and identifying military targets in surveillance images. Each project description includes the application's domain and its intended purpose, providing a comprehensive overview of how AI can be applied across different fields.}
\footnotesize
\centering
\begin{tabular}{|l|l|p{15.5cm}|}
\hline
Part. & AI Exp. & Selected AI Application Idea \\
\hline \hline
P1 & 3 yrs & \scriptsize{[Manufacturing / Logistics] Predicting demand for manufactured food products at various store locations (for planning manufacturing and shipping of goods)} \\
P2 & 7 yrs & \scriptsize{[Finance / Banking] Classify loan applications into those that should be approved/declined, or need further review (to assist and speed up the loan application review process)} \\
P3 & 6 yrs & \scriptsize{[Education / HR] Predict student dropout at universities (to proactively provide support to
those most vulnerable)} \\
P4 & 4 yrs & \scriptsize{[Manufacturing / Logistics] Predicting demand for manufactured food products at various store locations (for planning manufacturing and shipping of goods)} \\
P5 & 4 yrs & \scriptsize{[Art / Cartoon] Generate detailed sketches of cartoon cuts in the artist’s style based on the artist’s rough sketch (to help artists with the exploration of the design space)} \\
P6 & 2 yrs & \scriptsize{[Education / Recommendation] Providing personalized recommendations of courses (to help learners continue their path of learning on an online educational platform)} \\
P7 & 4 yrs & \scriptsize{[Manufacturing / Quality Control] Detecting defective pixels in newly manufactured televisions (to automate the process of items with defects)} \\
P8 & 1 yr & \scriptsize{\textls[-90]{[Productivity / Customer Support] Building an automated phone answering system that can respond to various consumer questions (to help reduce the burden of smaller companies that cannot keep its own customer care team)}} \\
P9 & 5 yrs & \scriptsize{\textls[-90]{[Productivity / Customer Support] Building an automated phone answering system that can respond to various consumer questions (to help reduce the burden of smaller companies that cannot keep its own customer care team)}} \\
P10 & 6 yrs & \scriptsize{[Rendering / Images] Manipulate the light source (direction, color, brightness, etc.) in an image (so that artists can remaster their images without needing to recapture images)} \\
P11 & 6 yrs & \scriptsize{[Military / Surveillance] Identifying key military activities and targets from surveillance images (to help generals quickly and accurately make decisions)} \\
P12 & 10 yrs & \scriptsize{[Art / Cartoon] Generate detailed sketches of cartoon cuts in the artist’s style based on the artist’s rough sketch (to help artists with the exploration of the design space)}\\
\hline
\end{tabular}
\label{tab:ai-experts}
\end{table*}

For the study, we recruited AI experts who have prior experience participating in AI application development with clients so that we can realistically simulate the information needs of the AI experts.
Because of our focus on the planning stage, we required that the participants have participated in the planning stage, directly interacting with their clients.

To obtain diverse perspectives, we recruited the participants through two channels: the authors' industry connections and referrals, and Upwork~\cite{upwork2024}.
For participants we recruited through our industry connections and referrals, we targeted AI engineers in AI solutions companies or non-AI companies since the AI engineers in these companies tend to have experience directly interacting with their clients during planning.
For participants we recruited through Upwork, we carefully reviewed their portfolios and prior experience published on their information pages to ensure that they have prior experience in building AI applications.
Through the recruitment process, we recruited a total of 12 participants (Table~\ref{tab:ai-experts}), from various countries, including the US, Egypt, India, South Korea, and Spain.

The study lasted 60-75 minutes and we compensated the participants with an equivalent of 50 USD or the rate at the hourly rate posted on Upwork, up to 50 USD per hour.

\subsection{Procedure}
\change{To separate out and examine the AI expert's information needs from the client, we recruited AI experts and had them go through simulated asynchronous discussions based on a compiled list of AI application ideas from the real world.
We had them organize their AI application development approaches for the execution stage by filling in an \textit{execution preparation document} that we compiled based on a model card format~\cite{mitchell2019model}, while noting any questions they need to ask the hypothetical client.
Based on the questions, we utilized affinity diagramming~\cite{kawakita1982original,kawakita1986kj} to arrive at a taxonomy of AI experts' information needs from the client.}

\subsubsection{AI Application Ideas List Compilation}
For a realistic setup for the participants of the study, we generated a list of AI application ideas based on the AI applications mentioned in the interviews in Section~\ref{sec:interview}, exploratory discussions with various clients and AI experts prior to the interview, and pilot studies.
Because the purpose of the study is to elicit information needs from the AI experts, we carefully limited the information provided was in the format of: ``[AI application description] ([purpose of the AI application])'' (Table~\ref{tab:ai-experts} right column).

\subsubsection{Execution Preparation Document Generation}
As a way to clearly identify the reasons why the AI expert would have specific information needs and to avoid the AI expert accidentally leaving out certain execution-related information during the limited duration of the study, we built an \textit{execution preparation document}.  
The execution preparation document is designed to capture the list of information that is directly involved in the AI application development.
It is designed so that filling in the document would require roughly the same amount of information as sketching out a technical approach in the execution preparation phase.
Specifically, the execution document includes not only information that is derived directly from the client's needs and expectations (e.g., target users, use cases) but also decisions and deductions made by the AI team based on the provided information (e.g., AI model to use).

\change{We generated this document based on \textit{model cards}, document forms used to communicate key information about an AI application to broad audiences, because model cards are already widely adopted across various AI expert communities ~\cite{google2024model,hugging2024model,kaggle2024model}.
Of the model card formats, we started with the model card format presented by Mitchell et al.~\cite{mitchell2019model}, as it has been reviewed through a peer-review process and not tuned to specific communities.}
However, a model card is aimed at disseminating an AI application after its development is complete, while the execution preparation document is aimed at organizing information required for execution preparation.
To amend these differences, we added missing fields related to the execution preparation phase to the model card and removed those that were irrelevant.
To identify the missing fields, we reviewed the contents mentioned in the interview and marked any contents that the interviewee mentioned as relevant to the execution preparation phase.
Next, we asked three AI experts within our organizations to fill out the model card based on the AI application development project they are participating in, while noting down any missing fields or irrelevant fields.
Some notable fields marked as missing and added in the final execution preparation document were about the timeline and the regulations and policies around data and AI use.
On the other hand, some fields marked as extraneous and removed from the final execution preparation document were model version and software license.

We ported the execution preparation document to Google Docs~\cite{google2024docs} for the study so that we can observe the participant's edits as they perform the task and so that we can use the comment feature for collecting their information needs.

\subsubsection{Study Execution}
Since the participants of the study were from various geographical locations, we performed the study on Zoom~\cite{zoom2022online} and Google Meet~\cite{google2024meet}, depending on participant preference.

After introducing the research team and the purpose of the study, we began by asking about the participant's background: amount of experience with AI and the project domains of some of their noteworthy AI application development projects.

Based on our findings from the interview that a client seeks an AI expert who is likely to have prior experience and expertise in their domain, we matched each participant with projects from the AI application ideas list that they are the most familiar with.
Specifically, we asked the participants to choose up to three project ideas from the AI application ideas list based on (1) their familiarity with the project domain and (2) the machine learning tasks potentially involved in approaching the project (e.g., object detection, question answering).
If the participant stated that their familiarity with a certain project idea greatly exceeds that with others, we assigned them to the project given that the same project was selected by at most one other participant.
Otherwise, we selected the project within their top 3 selection that no other participant had selected.

After deciding on the AI application project, we asked the participant to imagine a scenario in which a client has approached them to develop an AI application for them, and pause the scenario right after the client has only had a chance to state what the AI application would be about and the purpose of the AI application as given in the description.
As a method of collecting information needs of the AI experts, we asked the participant to fill in the execution preparation document while noting anything they either expect to hear from the client upon unpausing the scene or questions they would ask the clients as Google Doc~\cite{google2024docs} comments on the fields that require that information.
When filling in the study preparation document, we first asked the participants to skim through the fields to familiarize themselves with the contents of the document and then fill it in in the order they wish to.
We also instructed our participants to assume that their client has AI knowledge so that the participant would not limit themselves from listing out information needs due to their assumptions on what their client can answer.

Once the participant finished filling in the execution preparation document, we first checked whether they could imagine an initial execution plan if they were given answers to all the information needs they listed, 
Then, we asked them how the information needs would have differed if the client did not have sufficient knowledge about AI or if the AI application project was instead in a domain that they were not familiar with.
Finally, we asked them to freely express any other thoughts they had about the study.

\subsubsection{Result Analysis \& Taxonomization}

For analysis of the results, we first extracted a total of 227 instances of information needs from the Google Doc comments left on the execution preparation document.
Two of the authors put the instances of information needs into Miro~\cite{miro2024} and used an affinity diagram~\cite{kawakita1982original, kawakita1986kj} to come up with the lowest-level classes of information needs.
From these lowest-level classes of information needs, we repeatedly used affinity diagramming to arrive at higher levels of classes until no longer being able to further group similar classes of information needs.

\subsection{RQ1: Information Flow Taxonomy}
\begin{figure*}
    \centering
    \includegraphics[width=\textwidth]{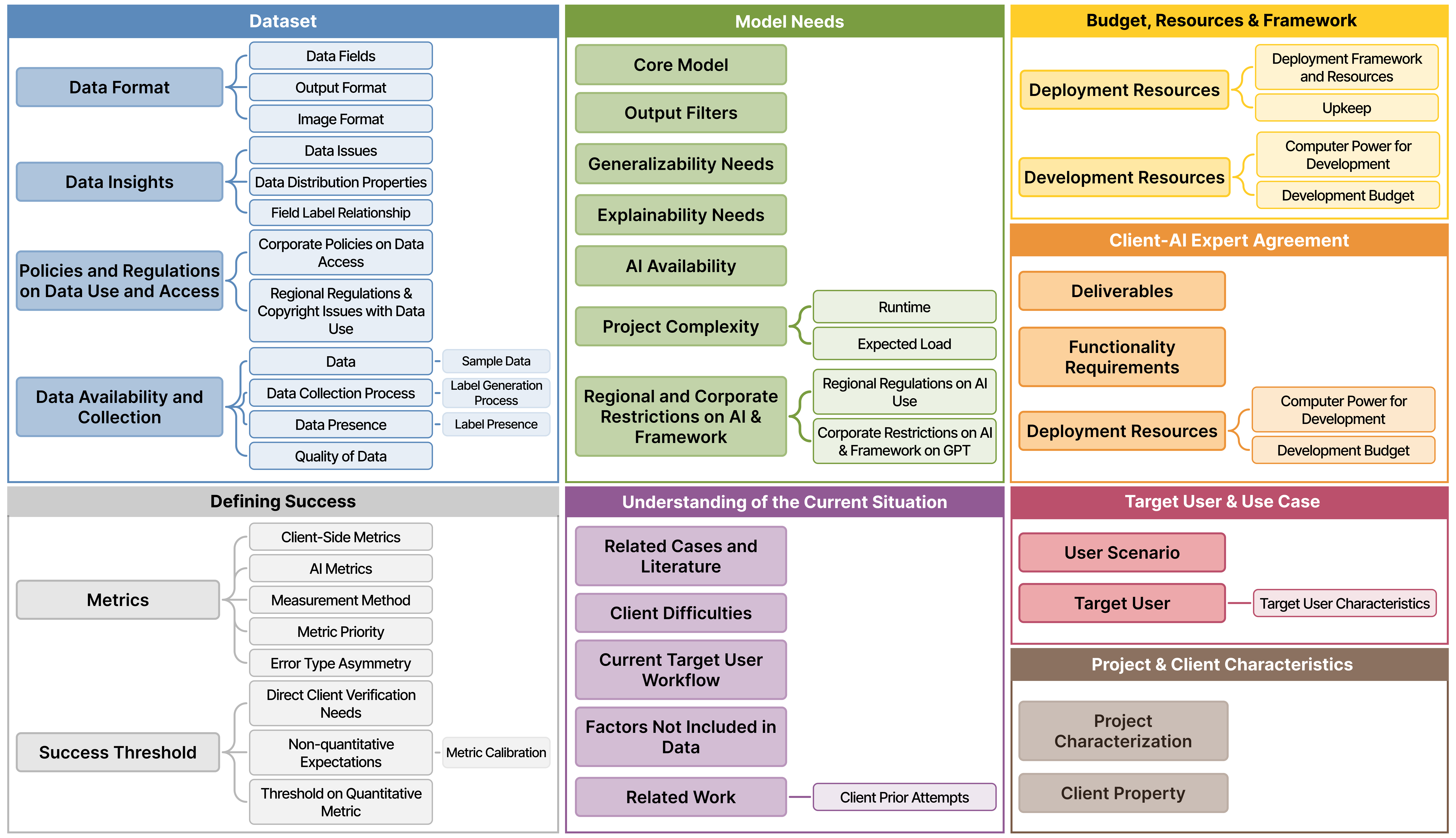}
    \caption{A taxonomy of the information needs of an AI expert from their client as they sketch out their technical approach.}
    \Description{This figure displays a multi-colored taxonomy of the information needs of an AI expert from their client. It is organized into three main columns and multiple rows, with the first two columns divided into two rows each, and the third column divided into four rows. From left to right, the first two columns are titled 'Dataset', 'Model Needs', 'Defining Success', and 'Understanding the Current Situation'. The third column, on the rightmost part of the diagram, is segmented into four rows titled 'Budget, Resources and Framework', 'Client-AI Expert Agreement', 'Target User and Use Case', and 'Project and Client Characteristics'.
    Starting from the top left, the 'Dataset' column, shaded in blue, includes subcategories such as 'Data Format', 'Data Insights', 'Policies and Regulations on Data Use and Access', and 'Data Availability and Collection', each with further detailed items.
    The next column, 'Model Needs', shaded in green, lists 'Core Model', 'Output Filters', 'Generalizability Needs', 'Explainability Needs', 'AI Availability', 'Project Complexity', and 'Regional and Corporate Restrictions on AI \& Framework', breaking down into more specific elements like 'Runtime' and 'Expected Load'.
    The 'Defining Success' column, shaded in gray, encompasses 'Metrics' and 'Success Threshold', with specifics such as 'AI Metrics', 'Measurement Method', 'Error Type Asymmetry', and 'Client-Side Metrics'.
    The 'Understanding the Current Situation' section, shaded in purple, includes 'Related Cases and Literature', 'Client Difficulties', 'Current Target User Workflow', 'Factors Not Included in Data', and 'Related Work'.
    The 'Budget, Resources and Framework' section, colored yellow, includes 'Deployment Resources' and 'Development Resources'.
    Next, colored in orange, the 'Client-AI Expert Agreement' includes 'Deliverables', 'Functionality Requirements', and 'Deployment Resources'.
    'Target User and Use Case', colored in red, includes 'User Scenario' and 'Target User'.
    Lastly, the 'Project and Client Characteristics' section, colored in brown, includes 'Project Characterization' and 'Client Property'.}
    \label{fig:taxonomy}
\end{figure*}

As a result of our analysis, we obtained a taxonomy of the information needs of AI experts from their clients (Figure~\ref{fig:taxonomy}).
The taxonomy includes a total of 71 nodes and has a maximum depth of 4.
We provide a brief overview of the top-level nodes of the taxonomy in the order of the number of instances of information needs corresponding to each top-level node.
Please refer to Figure~\ref{fig:taxonomy} for further details.
\begin{itemize}[leftmargin=*]
    \item \textbf{\textit{Dataset (88 instances; 39\%)}} includes information about the data format (input \& output), availability of the data and the collection method, domain insights into the data, and regulations and restrictions around data use; this information influences various parts of the AI expert's decisions, including the AI model to use, need for preprocessing or data augmentation, metrics, and feasibility assessment.
    \item \noindent\textbf{\textit{Model Needs (39 instances; 17\%)}} includes information about the client's needs in selecting the AI model to use (e.g., explainability, generalizability, runtime) as well as regulations on AI use; this information directly influences the AI expert's decision around the choice of the AI model to utilize.
    \item \noindent\textbf{\textit{Defining Success (30 instances; 13\%)}} includes information about the metrics for success and the threshold to meet on the metric for the application development to be satisfactory to the client; this information directly affects the AI metrics and target thresholds that the AI expert would use during development.
    \item \noindent\textbf{\textit{Budgets, Resources \& Framework (22 instances; 10\%)}} includes information on the available resources (e.g., GPU, API) and budgets available for development and deployment; this information serves as the limiting factor for the AI expert as they determine the AI model to use.
    \item \noindent\textbf{\textit{Understanding the Current Situation (17 instances; 7\%)}} includes information on the client's workflow and difficulties as well as related work and the client's prior attempts; this information helps the AI expert gain the required background knowledge and the information on related work and client's prior attempts in particular inform potential target thresholds.
    \item \noindent\textbf{\textit{Client-AI Agreement (14 instances; 6\%)}} includes information about the timeline, deliverables, and functionality requirements that the resulting AI application must meet; this information informs the AI expert about what terms they would expect to be bound to as a part of the client-AI expert collaboration.
    \item \noindent\textbf{\textit{Target User \& Use Case (10 instances; 4\%)}} includes information about who would be using the resulting AI application and the usage scenarios; this information serves as background knowledge for the AI expert as they reason about various decisions, including those about the input and the output.
    \item \noindent\textbf{\textit{Project \& Client Characteristics (7 instances; 3\%)}} includes information about the nature of the project (e.g., research prototype, global deployment, etc.) and the philosophy of the client company; this information mostly serves as a guide for the AI expert in forming expectations about the collaboration style.
\end{itemize}

\subsection{RQ2: Dimensions Affecting Information Needs}
We found four dimensions affecting the information needs of AI experts from their clients: (1) AI knowledge level of clients, (2) domain knowledge level of AI experts, (3) input data type, and (4) sensitivity of data handled in the domain. 
We discuss how the information needs depend on each of these dimensions and the implications each of the dependencies has on the documentation form design.

\subsubsection{AI Knowledge Level of Clients}
In the study, we asked the participants to assume that they possess AI expertise in order to elicit information needs from the participants without holding back questions based on the assumption that the client will not be able to answer them.
While it is true that many of the clients do possess some AI knowledge or learn about AI along the way, clients who do not possess AI knowledge are common (noted separately by P2, 7, 10-12).
In general, the participants stated that as AI experts, they would be much more likely to make the more technical decisions in defining success (AI metrics for development progression, threshold on the AI metric) on their own instead of asking the clients about the technical details (P5, 7, 9, 12).
In addition, they also stated that they would give \textit{directives} about the data requirements (e.g., format, quantity, collection methods) instead of expecting that the clients would be able to make the right decisions on their own (P2, 5, 7, 12). 
On the contrary, participants were more willing to invite the clients to determine the tasks that are typically considered the tasks of the AI expert, such as modeling methods, fine-tuning methods, or even AI-related coding (P3, 5), although such client intervention in the AI expert's traditional role is not always welcome (P11). 

\vspace{0.05in}
\noindent\textbf{\textit{Implications for Documentation Form Design.}} 
A generalizable documentation form should be usable by clients with varying degrees of AI knowledge.
The documentation form should allow more control of the technical decisions for clients who possess AI knowledge, while not requiring the same information from clients who do not.
One way to achieve this goal would be to include an optional input field about the technical decisions so that clients who have AI knowledge can express their voice, while clients without sufficient AI knowledge can simply skip the input field.

\subsubsection{Domain Knowledge Level of AI Experts}
While clients tend to choose AI experts who are more likely knowledgeable about the problem domain, some AI experts admitted that they had instances in which they had to work with clients in a problem domain they were unfamiliar with (P7, 10).
We believe that this is because of the existence of a long tail when it comes to the space of all possible AI application domains (e.g., P7's prior experience with tattoos).
To understand how information needs would change in these cases, we asked the hypothetical question about what would change about the information needs if the participant were instead given a problem in an unfamiliar domain.
Participants stated that they would inquire much more heavily about understanding the target user workflow (P1, 4, 6, 7, 9, 10) and insights into the data and the data fields (P1, 5, 8, 11).
Yet, many participants said that they would also seek information sources other than the clients for additional information (P2, 7, 9-12).

\vspace{0.05in}
\noindent\textbf{\textit{Implications for Documentation Form Design.}} 
A generalizable documentation form, when filled out, should contain sufficient domain information for the AI experts to comprehend.
Because the amount of information needs increases with a decrease in an AI expert's domain expertise, the documentation form should be made for the AI experts with the least domain expertise in order to be usable by any AI expert.
Hence, the documentation form should be inclusive of the information needs of AI experts without domain expertise.

\subsubsection{Input Data Type}
\label{sec:input-data-type}
The data the AI application takes as input can take many different forms (e.g., text, images, tabular data).
We observed that there are information needs specific to certain data formats.
For instance, if the input is of an image format, some of the commonly mentioned information needs were around the dimensions and resolution of the image (P7, 11, 12) and the number of color channels in the dataset (P11, 12).
In comparison, if the input is of a tabular format, information needs focus more on the fields of the data (P1, 3, 4).

\vspace{0.05in}
\noindent\textbf{\textit{Implications for Documentation Form Design.}}
Unlike the knowledge levels discussed in the previous sections, the input data type is a categorical variable.
Thus, it is necessary to identify possible data types the input data can take and understand what can be asked in common across multiple input types and what needs to be asked differently for each input type.
Input data can be multimodal and include multiple data types (e.g., image question-answering, data tables containing images), but attempting to deal with each data type in separation could lead to exponential growth in the cases that the documentation form needs to cover.
Therefore, it would be more tractable to break down each multimodal input type into each modality and deal with them in isolation.
Furthermore, there exists a long tail of input data types (e.g., 3D mesh data) and may be a place for adaptations for specific domains instead of trying to deal with every possible input data type; domains dealing with less common input data types can extend the documentation form based on their needs.

\subsubsection{Sensitivity of Data Handled in the Domain}
Another dimension affecting information needs is the sensitivity of the data being handled in the domain.
While information needs on client policies on data sharing hold for all domains, information needs on regulations are more pronounced in some domains than others.
In our study, we observed stronger information needs when the AI expert needs to deal with educational data (P3, P6), military data (P11, mentioned by P10 in their previous experience), copyrighted artwork (P5, 12), medical data (mentioned by P11 in their previous experience), and personal finance data (P2). 

\vspace{0.05in}
\noindent\textbf{\textit{Implications for Documentation Form Design.}} 
Since there are strictly more information needs if the data being handled in the domain is sensitive, the documentation format can by default ask about the regulations and copyright issues regarding the data.

%% file: sections/05-workbook.tex
\section{The \wbname{} Workbook: Design and Case Studies}
\label{sec:workbook}
\begin{figure}
    \centering
    \begin{subfigure}{\linewidth}
        \includegraphics[width=\textwidth]{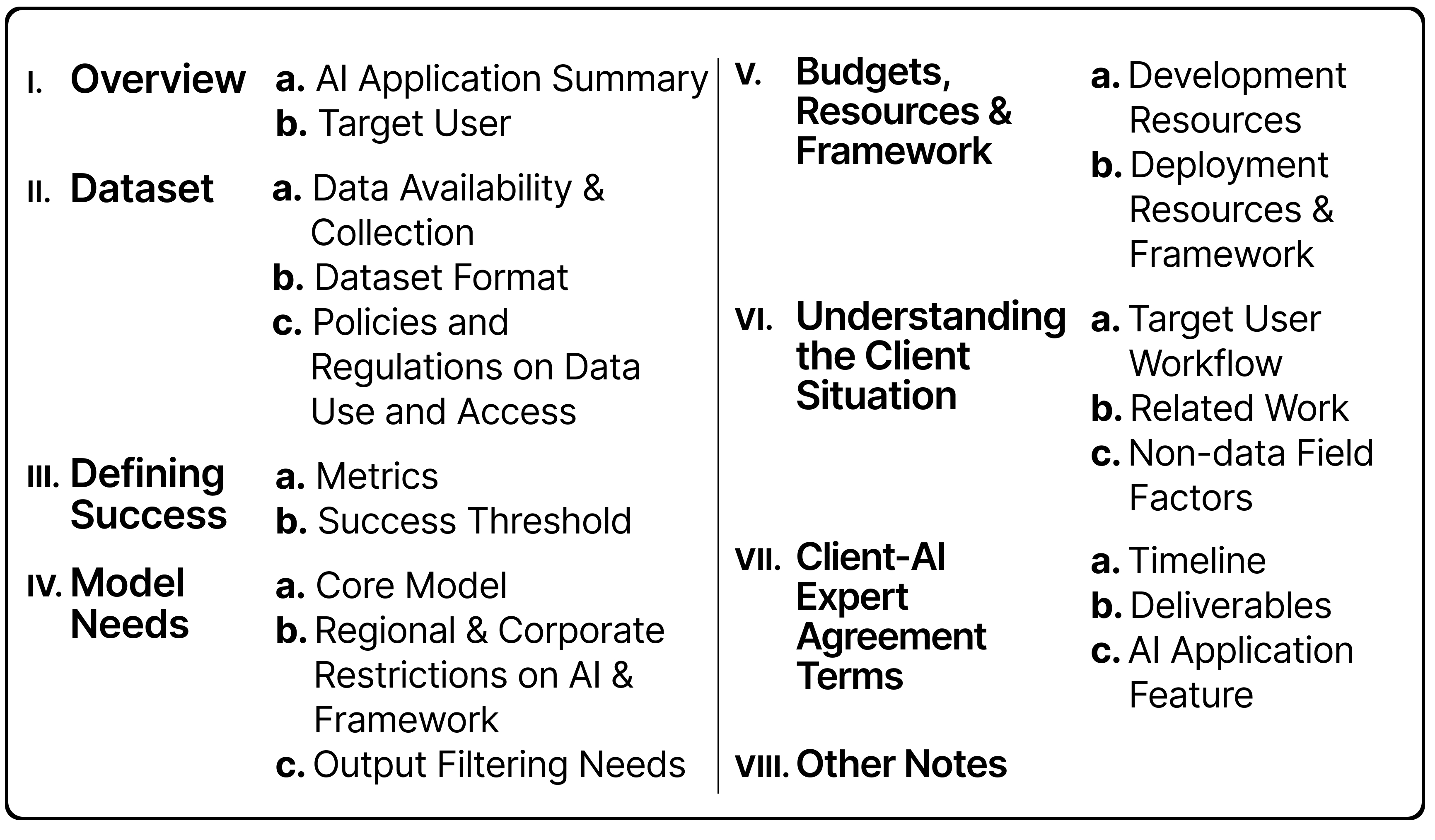}
        \caption{Chapters \& subchapters.}
        \Description{(a) presents the 'Table of Contents' of the workbook, detailing the chapters and subchapters. It lists the following sections: I. Overview, II. Dataset, III. Defining Success, IV. Model Needs, V. Budgets, Resources \& Framework, VI. Understanding the Client Situation, VII. Client-AI Expert Agreement Terms, and VIII. Other Notes, each with specific subsections.}
    \end{subfigure}
    \begin{subfigure}{\linewidth}
        \includegraphics[width=\textwidth]{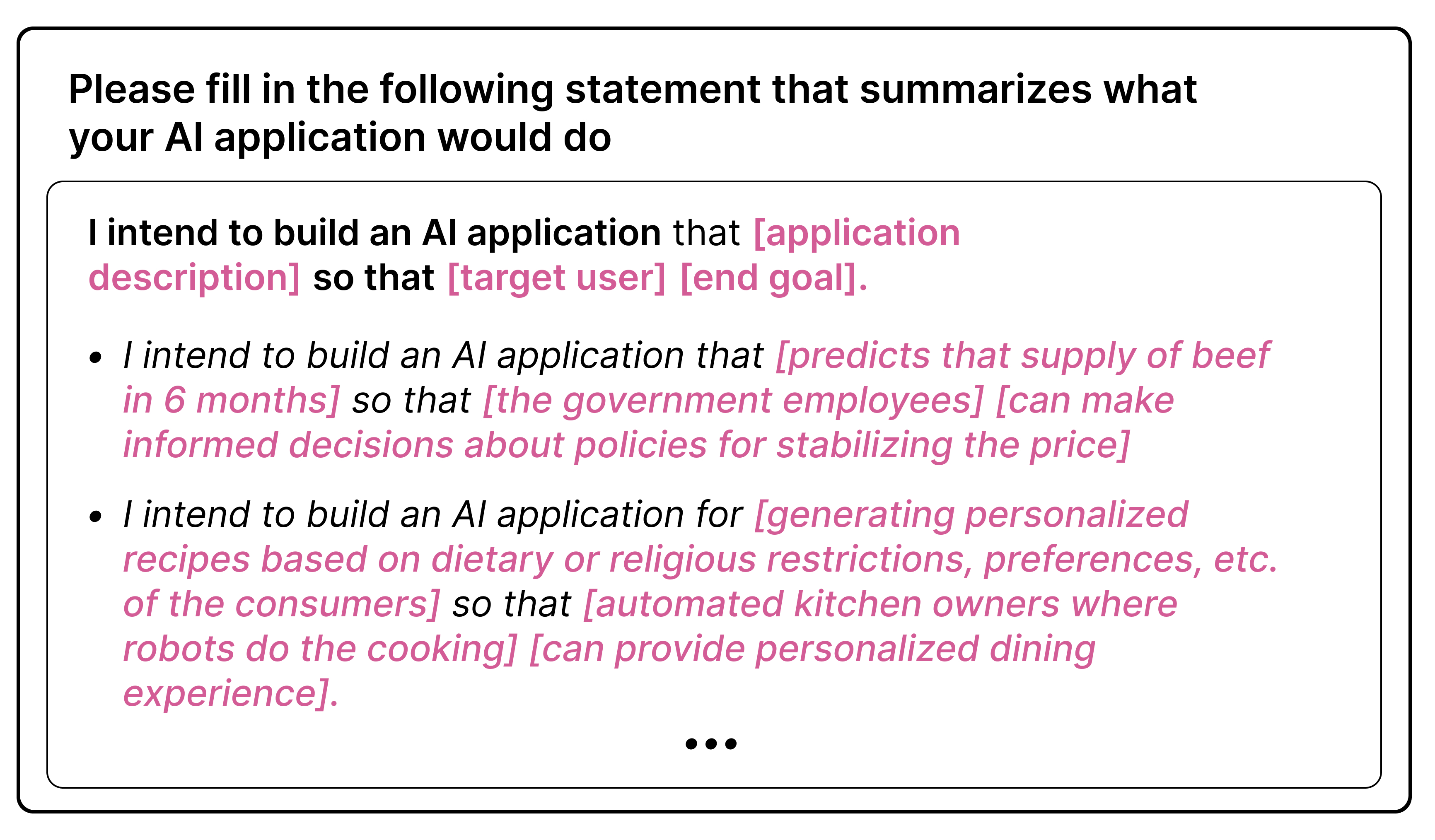}
        \caption{Example question.}
        \Description{(b) shows an 'Example Question' from the workbook. This section prompts users to complete a template sentence summarizing what their AI application would accomplish. The sentence is structured as 'I intend to build an AI application that [application description] so that [target user] [end goal]'. Below this template, two examples are provided: one about predicting the supply of beef for government employees to inform policy decisions, and another about generating personalized recipes based on dietary restrictions for automated kitchens to enhance the dining experience.}
        \end{subfigure}
    \caption{An outline of the workbook contents and a sample of the workbook content style.}
    \label{fig:workbook}
\end{figure}
Based on the results of the \change{formative} interview (Section~\ref{sec:interview}) and the main study (Section~\ref{sec:experiment}), we built the \wbname{} workbook that outlines the information needs of the client to help support effective collaboration between AI experts and clients.
In this section, we describe the design of the workbook as well as the two case studies we conducted to understand the values of the workbook in the use cases we identified through the interview in Section~\ref{sec:interview}.

\subsection{Workbook Design}
In designing the workbook, we had the goal of supporting the two use cases we discovered through the \change{formative} interview:

\vspace{0.025in}
\noindent\textbf{\textit{G1.}} The workbook should serve as an effective guide for clients as they prepare information for collaboration.

\vspace{0.025in}
\noindent\textbf{\textit{G2.}} The workbook should serve as an effective discussion guide and checklist for AI experts as they elicit information from their clients.

We selected Google Docs~\cite{google2024docs} as the platform for hosting the workbook because it not only includes an outline of the chapters and subchapters that helps with easy navigation, but also allows easy collaborative edits and asynchronous communication through comments.
However, the design of the workbook itself does not depend on the platform and can be ported to other platforms depending on user needs.

We followed the 5-step process outlined below to build the workbook:

\vspace{0.05in}
\noindent\textit{\textbf{Step 1. Choosing a model documentation form.}}
We first reviewed various structured documentation forms (e.g., questionnaires, workbooks) within and outside of the AI application development domain to find a model format on which we can base the design of our documentation form.
After rounds of internal discussions and tests on the information communication capacities of the different formats, we decided that the \textit{workbook} format utilized by the Google People + AI guidebook~\cite{google2019people} (example chapter worksheet: \cite{google2019user}), a guidebook that introduces best practices and examples for designing with AI, is a good model documentation form to start with; it is a documentation form also related to AI application development that has already been adopted in actual worksites~\cite{yildirim2023investigating}.

\vspace{0.05in}
\noindent\textit{\textbf{Step 2. Structuring the workbook into chapters \& subchapters.}}
The decision on the holistic structure of the workbook came from the model workbook.
We first took the notion of chapters into our workbook by translating the top-level nodes in the information flow taxonomy into chapters.
Because of the size and complexity of our taxonomy, we further subdivided each of the chapters into subchapters for navigability, which resulted in a total of 8 chapters and 19 subchapters (Figure~\ref{fig:workbook}a).

\vspace{0.05in}
\noindent\textit{\textbf{Step 3. Generating question contents.}}
Afterward, we generated the question contents by looking at the taxonomy as well as each instance of information needs we used to build the taxonomy.
Whenever possible, we extracted the wordings that the AI experts used to express their information needs to the clients during the study in Section~\ref{sec:experiment} to inform the exact wordings of the questions.
For the style of the questions, we also adopted the style of the questions used in the workbook: fill-in tables, fill-in-the-blanks, and free-form questions.

\vspace{0.05in}
\noindent\textit{\textbf{Step 4. Constructing examples.}}
Along the lines of G1, to further help clients, who may not have much AI expertise, comprehend and perform tasks around the workbook, we added examples or additional hints about filling it in using lighter text (e.g., Figure~\ref{fig:workbook}b).
We constructed the examples by going through the AI application ideas list used in the study in Section~\ref{sec:experiment} and trying to answer the questions ourselves and selecting the most representative samples.

\vspace{0.05in}
\noindent\textit{\textbf{Step 5. Iterating on wording and examples.}}
We reflected on the suggestions for changes in wording and examples from three test runs with domain experts within our organizations.

\vspace{0.05in}
We include the complete \wbname{} workbook in the supplemental materials.

\subsection{Case Studies: Application in the Real World}
\begin{table}
    \caption{A summary of the AI application development projects covered in the two case studies.}
    \Description{Table summarizing AI application development projects from two case studies focused on the use of a workbook. The table is divided into two sections. The first section, 'Case Study 1: Worksheet as a Client Guide’, lists projects CS1A through CS1E with domains including Medical Imaging, Marketing and Consumer, Oncology and Hematology, Digital Fashion and App Design, and Medical Robotics. Each project entry details the years of experience of the AI expert and the client in their respective domains, ranging from 3 to 11 years for AI experts and 2 to 40 years for clients. The second section, 'Case Study 2: Worksheet as a Reflection Guide for AI Experts’, includes a single project, CS2 in the Telecommunication domain, with the AI expert having 8 years of experience and the client's domain experience undisclosed. This table provides a detailed overview of the diverse expertise involved in the case studies and the different project domains addressed.}
    \centering
    {\footnotesize
    \begin{tabular}{|l|l|l|l|}
        \hline
         Project & Project Domain & AI Expert Exp. & Client Domain Exp. \\
         \hline\hline
         \multicolumn{4}{|l|}{Case Study 1: Worksheet as a Client Guide} \\
         \hline
         CS1A & Medical Imaging & 3 yrs & 6 yrs \\
         CS1B & Marketing / Consumer & 3 yrs & 5 yrs \\
         CS1C & Oncology / Hematology & 11 yrs & 40 yrs \\
         CS1D & Digital Fashion / App Design &  4 yrs & 3 yrs \\
         CS1E & Medical Robotics & 4 yrs & 2 yrs \\
         \hline\hline
         \multicolumn{4}{|l|}{Case Study 2: Worksheet as a Reflection Guide for AI Experts} \\
         \hline
         CS2 & Telecommunication & 8 yrs & Undisclosed \\
         \hline
    \end{tabular}
    }
    \label{tab:case-study}
\end{table}

To gain initial insights into whether we met our goals (G1, G2) for supporting the use cases identified in Section~\ref{sec:interview}, we performed case studies based on the two use cases. 

In addition to the two use cases, the AINeedsPlanner workbook has been adopted by \change{AlgorithmLabs}, an AI solutions company located in \change{South Korea}.
However, due to the sensitivity of the information as well as the legal issues around non-disclosure agreements, we are unable to publicly share further detail.

\subsubsection{Case Study 1: Workbook as a Client Guide}
\change{To understand the value of the workbook as a guide for the client as they determine information in the collaboration preparation phase, we specifically recruited five domain experts with clear needs for AI application development to fill in \wbname{} over one week. We shared the filled workbooks with AI experts for initial feedback to gain insights about how this leads into the following main discussion phase.}

\vspace{0.05in}
\noindent\textbf{\textit{Participants.}}
In order to avoid the restrictions arising from the non-disclosure agreements between AI solutions companies and their clients, we recruited the domain experts ourselves and acted as a mediator between them and the AI solutions companies.
We reached out to \change{five} domain experts who have clear needs to make use of AI applications through referrals through our professional and personal connections (Table~\ref{tab:case-study} Block 1).
\change{The five domain experts were in different domains and had varying levels of domain expertise. Specifically, while the domain expert recruited for CS1A, CS1C, and CS1D had very little or no prior experience with AI, the domain expert recruited for CS1B and CS1E reported that they possess some prior experience with AI.}

\vspace{0.05in}
\noindent\textbf{\textit{Procedure.}}
We first met with each of the domain experts to explain the process, either on the phone or in person.
During this meeting, we asked briefly about their AI application development needs, their experience in the domain, and their experience with AI.
We asked them to assume that they would be working with an AI expert for the development and gave them the \wbname{} workbook that will be sent to the AI expert to begin an initial discussion.
After answering any questions from the domain experts, we asked them to fill in the workbook and return it to us with a completed short survey about their experience within one week.
Once we received the completed workbooks from the domain experts, we passed them to an AI expert recruited from an AI solutions company for review and asked for comments.

Because the AI application projects are to be developed, the domain experts asked us to keep the details of the plan private for publication.

\vspace{0.05in}
\noindent\textbf{\textit{Results.}}
\change{All} domain experts were able to fill in the workbook without much difficulty, including \change{the domain experts with little or no experience in AI (CS1A, CS1C, CS1D), with the domain expert in CS1A stating} that ``\textit{I was concerned that filling in the form would require knowledge of AI, but that was not the case.}'' 
This result suggests that our workbook may be able to support even the clients without much AI knowledge.
\change{In particular, the domain experts of CS1A, CS1C, and CS1E specifically pointed out that the examples were especially helpful in knowing what to fill in for each question.}

\change{
Moreover, the domain experts indicated that the workbook successfully guided them as they considered and gathered information about the AI application (CS1C, CS1D, CS1E). For example, the domain expert in CS1E stated that the workbook breaks down the various aspects of AI application planning beyond what they would have thought about on their own. In addition, the domain expert in CS1C reported that the questions of the workbook helped them actively seek specific information from a collaborator in a different institute regarding their successful introduction of a similar AI application. They reported that this helped them further develop the details of their AI application idea.
In fact, the AI expert for CS1C stated that the detailed information provided in the workbook is well beyond what they would expect to be able to elicit from less prepared clients with multiple extended discussion sessions.
}

\change{Based on the AI expert feedback, while the filled workbooks from client-side preparation had varying levels of imperfections, the AI experts were able to comprehend the key ideas and identify components that require further discussion.
For example, in CS1B, the AI expert was able to pinpoint that they will need to obtain further discussions on the data label format and the purpose of the AI application, and that some calibrations will be needed about the domain expert's model needs. 
As another example, in CS1D, the AI expert expressed specific needs to further understand the current workflow in the domain as well as discuss the quantitative target measurements.
}

\change{These results suggest that the value of the workbook arises not from the fact that it is able to draw a complete AI application from the client, but from the fact that it guides clients to think about and develop AI application ideas during the collaboration preparation phase so that the AI experts can concretely reason about the plan and recognize additional discussion points.}
This could reduce the amount of back-and-forth communication and hence \change{pave the way for efficient communication between clients and AI experts during the main discussion phase.}

\subsubsection{Case Study 2: Workbook as a Communication \& Reflection Guide for AI Experts}
The second case study is aimed at understanding the value of the workbook as a guide for the AI experts as they communicate with the client in the main discussion phase.

\vspace{0.05in}
\noindent\textbf{\textit{Participants \& Procedure.}}
For the second case study, we provided our workbook to an AI engineer at an AI solutions company and asked them to utilize the workbook while performing planning in their next AI application development project in any way they wish to (Table~\ref{tab:case-study} Block 2).
After the planning was complete, the AI engineer reported to us about their experience.

Both the client company and the AI engineer agreed to the AI engineer sharing their experience for academic publication as long as the contents of the project and the identities of the stakeholders are concealed.

\vspace{0.05in}
\noindent\textbf{\textit{Results.}}
The AI engineer reported to us that they used the workbook to guide communication with the client and record the discussions.

The key value of the workbook for the AI engineer was as a tool for ensuring completeness of the communication. 
The AI expert specifically noted that they appreciated the completeness of the document, and especially the section on budgets, resources \& framework. 
This is because in their prior experience, they forgot to discuss budgets and resources during the planning phase and this caused confusion later on when they discovered that the client did not have sufficient resources to deploy the model.

%% file: sections/06-discussion.tex
\section{Discussion}

The two case studies illustrate the potential of using \wbname{} for various purposes in the clients-AI experts collaboration process. In this section, we discuss the broader utility of the \wbname{} workbook, incorporating more design attributes and stakeholders in the workbook by comparing it with a prior taxonomy and a potential for systematic support.

\subsection{Broader Utility of the Workbook for Clients and AI Experts}

Our case studies showed that \wbname{} was able to elicit key information for building AI applications from clients, give them an awareness of the information they need to prepare before meeting with AI experts and help AI experts reflect on the information needed from the clients. \wbname{} offers an easy and effective planning experience by employing the ``Recognition Rather Than Recall'' strategy~\cite{nielsen10usability}, a useful design pattern for interface design. The ease of use of the workbook could encourage more active participation of the clients in the application-building process, contributing to better contextualized AI applications~\cite{sloane2022participation}. 

It also illustrates the potential for \wbname{} as a learning material for the clients to understand what needs to be prepared for building AI applications. The workbook not only provides essential questions but also gives contextualized instructions (or even lectures) about the corresponding AI concepts for each section so that the clients can answer the questions in an informed way, reducing the technical knowledge gap between the clients and AI experts. On the flip side, AI experts can also benefit from having learning material for understanding the clients' domain contexts, reducing the domain-specific knowledge gap between the two. 

Additionally, \wbname{} served as a boundary object facilitating shared understanding between the client and AI experts by clearly scoping, organizing, and retaining the key information needed in the planning stage. An interesting note from CS2 was that the workbook can promote more structured discussion with clearer sub-stages in the planning. For example, some of the sections in the workbook (e.g., metrics, success thresholds) need a discussion between the two, but other sections (e.g., detailed dataset information, available budgets, and resources) mostly require information from the clients. Therefore, a possible planning structure would be first making a clear consensus about the information needed from the clients and then having a discussion for making decisions that require both stakeholders' inputs.

\subsection{Extending Design Attributes and Stakeholders in the Workbook}

\begin{figure}
    \centering
    \includegraphics[width=\linewidth]{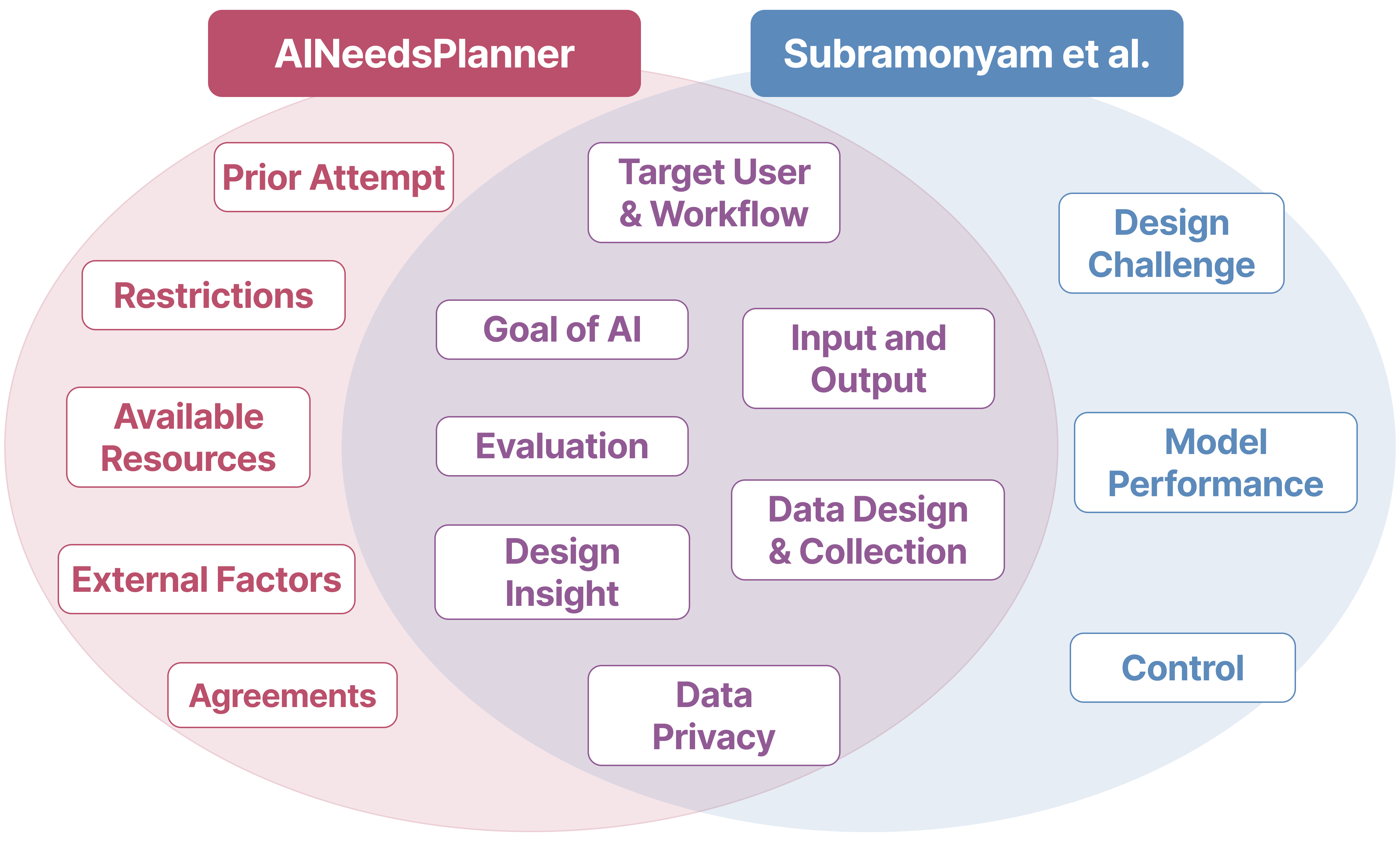}
    \caption{A Venn diagram showing the commonalities and differences between the information covered by \wbname{} and the information covered by Subramonyam et al.~\cite{subramonyam2022solving}.}
    \Description{The figure displays a Venn diagram with two overlapping circles on a white background. The circle on the left, shaded light pink and labeled 'AINeedsPlanner', lists factors such as 'Prior Attempts', 'Restrictions', 'Available Resources', 'External Factors', and 'Agreements'. The circle on the right, shaded light blue and labeled 'Subramonyam et al.', includes 'Design Challenge', 'Model Performance', and 'Control'. The overlapping section in the middle, which represents common factors considered by both, contains 'Target User \& Workflow', 'Goal of AI', 'Evaluation', 'Design Insight', 'Input and Output', 'Data Design \& Collection', and 'Data Privacy'. This diagram visually illustrates the shared and unique considerations between the two approaches or entities, AINeedsPlanner and Subramonyam et al.}
    \label{fig:taxonomy-comparison}
\end{figure}

\wbname{} focused on the planning stage between clients and AI experts, identifying key information needed in the planning stage. The workbook can be extended by incorporating more stakeholders as well as design attributes of AI applications.

To understand the difference between \wbname{} and known design attributes of AI applications, we compared the \wbname{} workbook with a prior taxonomy of Human-AI interaction guidelines~\cite{subramonyam2022solving} by listing dimensions of each taxonomy and grouping similar dimensions. Figure~\ref{fig:taxonomy-comparison} shows a Venn diagram presenting the relationship between the dimensions of each taxonomy. Out of 15 dimensions of the prior taxonomy, \wbname{} covers 9 dimensions (4 training data, 2 AI models, 2 AI-powered user interface, and 1 human mental models categories from the prior taxonomy) whereas 6 dimensions (2 human mental models, 2 AI-powered user interface, and 2 AI models categories) were not the core focus of the workbook. We found three themes of the 6 dimensions (Design Challenge, Control, and Model performance) and hypothesized possible reasons for not having them as the core focus in \wbname{}. 

Design challenge discusses how to shape users' mental model and interaction model, which is one of the important considerations in designing usable AI applications~\cite{amershi2019guidelines, google2019people}. However, such design aspects were not brought up during the planning stage as the focus was mostly on the technical side of AI applications rather than user experiences. Depending on how important it is to design usable AI applications in the project, such design aspects can be covered in the planning potentially by incorporating designers as another stakeholder and prompting the other stakeholders to consider such design aspects through the workbook. 

The control aspect was not emphasized in the planning process as well. A possible reason could be a lack of AI experiences for the clients; clients might not have needs for controlling AI as they would not expect inconsistent and imperfect behavior of AI~\cite{dzindolet2003role}. Since it is difficult to imagine AI behavior in the planning stage~\cite{yang2020re, hong2021planning}, prototyping AI behavior~\cite{subramonyam2021protoai} even in the planning stage could facilitate the discussion of AI control. 

Finally, \wbname{} does not focus on how to ensure model performance for diverse use cases. A possible reason would be around the initial scope of the target AI application; clients would be interested in building AI applications for their specific contexts and use cases first for feasibility checking, which can be expanded later once the application is successfully deployed and used. 

The planning workbook can incorporate these important design attributes as well, depending on the characteristics of clients' needs and contexts. It illustrates the need for extending the workbook that can be adapted to various stakeholder combinations. For example, having designers in the planning stage, as well as clients and AI experts, may need additional sets of questions, which helps create a more robust plan considering both technical and user experience in the early stage. The managerial role is another important stakeholder who has the power of decision-making as well as budget management, which might need more sets of detailed practical questions. As \wbname{} can be a good starting point, it can be an effective approach to start planning with \wbname{} in such diverse settings, come up with questions that are important in such specific context, and design another version of the workbook that can be more specialized in the contexts.

\subsection{Towards Designing Interactive Support for the Planning Process}

The AI application planning through a collaboration between clients and AI experts is a complicated process, which brings multiple challenges, such as eliciting information from clients who are not knowledgeable in AI, reducing the knowledge gap between clients and AI experts for the client contexts, and AI, and creating plans that align with the clients' AI needs. \wbname{} addresses such challenges as a boundary object that facilitates clear communications on the same page. \wbname{} is a static documentation format, but we can go beyond that by designing interactive systems that provide a more adaptive planning experience, considering the contexts of planning (e.g., types of AI applications, participating stakeholders, and knowledge gaps between stakeholders). 

As we noted in the prior discussion, the design attributes that need to be considered in the planning depend on the context of planning, such as client domain context and participating stakeholders. Based on important design attributes to be considered, the system can adaptively organize the questions, leveraging questions in \wbname{}. For example, designing AI applications for a high-stake scenario (e.g., medical domain~\cite{cai2019hello}) might need to put more emphasis on the explainability and controllability of AI with designers in the collaboration process. Then, the system might ask questions about existing practices in verifying prediction results. 

Eliciting information from the clients can also benefit from having interactive systems. Research showed that an automated chatbot can elicit quality responses compared to an online survey, while providing engaging experiences~\cite{xiao2020tell}. Asking follow-up questions~\cite{ge2022should} can be effective in getting detailed and well-aligned responses for the question. Scaffolding the responses via autocompletion by sharing client contexts (e.g., suggesting task-specific measures for the types of AI applications and suggesting potentially relevant features to the label when describing data formats) can reduce clients' efforts in the knowledge elicitation process. 

As a prototype is commonly used as a boundary object in software development~\cite{huber2020use}, supporting quick prototyping of AI applications based on the planning contents has the potential to significantly help stakeholders be on the same page. Automated techniques such as AutoML~\cite{he2021automl} and UI prototyping techniques~\cite{subramonyam2021protoai} can be employed, but it needs more investigations to create such prototypes based on the planning contents. Another promising technique to have a concrete artifact could be finding relevant projects in the wild (e.g., from Papers With Code~\cite{papers2024}) to assess the feasibility of the current plan.

%% file: sections/07-limitations-future-work.tex
\section{Limitations and Future Work}

\vspace{0.05in}
\noindent\textbf{\textit{Conducting more in-depth case studies in real world.}} Although our two case studies suggest that \wbname{} does support client-AI expert collaboration, the coverage of project domains was limited and for CS2, we only had limited access to just the AI expert's experience due to a non-disclosure agreement between the AI expert and the client.
Hence, we believe that more in-depth studies observing both sides of the client-AI expert collaboration in more cases would help us further our understanding about the value of our workbook.
\change{While \wbname{} is a workbook for generic AI applications and we have identified dependencies of information needs specific on certain data types (Section~\ref{sec:input-data-type}, we believe that the in-depth studies could lead to insights that can lead to more specialized workbooks tailored to specific application domains or application types.}
Future work can also study put our workbook to use in other types of collaborations, such as intra-team collaborations or collaborations with additional collaborators (e.g., UI/UX designers, managerial roles) and find ways to generalize our workbook.

\vspace{0.05in}
\change{
\noindent\textbf{\textit{Guiding assessments of risks and unintended consequences.}}
Assessing the risks and unintended consequences of AI applications has been widely recognized~\cite{vandepoel2023ai}. 
Some of the AI experts in the main study hinted at the need to understand the risks and consequences.
However, they considered them as components of other information needs, not as an independent information need on their own.
For example, P10 expressed information needs on potential harms of certain types of errors, but as a part of understanding and determining the metrics and the target values; P11 expressed information needs on the `seriousness' of the consequences of the project while trying to understand nature of the project.
Based on these results, we include a question about error type asymmetry within the section on defining success and a question about consequences of malfunction within the section on project and client characteristics.
Yet, we believe that the importance of discussions about risks and unintended consequences of AI applications will continue to rise with rapidly growing AI capabilities.
Hence, we suggest that future work further explores the information needs around risks and unintended consequences to form an independent section with additional emphasis.
We believe that this will steer discussions between AI experts and their clients towards responsible, ethical, and safe usage of AI.
}

\vspace{0.05in}
\noindent\textbf{\textit{Supporting input formats for the workbook beyond plain text.}}
The current form of \wbname{} only allows the users to express themselves using plain text.
While text is an expressive medium that can capture most if not all intents, it is not necessarily the most efficient medium for expressing, communicating, and recording information.
For example, while the workflow in the domain can be expressed in text, formats like a flow chart may be more effective in communicating the information.
To further support this, the AI expert in CS2 of our user studies further supported the idea and suggested that although they could fill in the workbook in the current format, they would have had an easier time filling the workbook if it had supported free-form sketches or diagrams.
Based on these observations, future work could either integrate tools that can accept input beyond plain text (e.g., Figma~\cite{figma2024}, Miro~\cite{miro2024}) or conversely attempt to build our workbook into these tools.

\vspace{0.05in}
\noindent\textbf{\textit{Adaptations to the shifting AI landscape.}}
The field of AI has been moving rapidly in recent years with frequent introduction of novel models and APIs (e.g., generative models).
Because it is a recent invention, we had only one sample of \textit{successfully concluded} projects around generative models in our interview in Section~\ref{sec:interview} (IT4).
On the contrary, many AI experts from the interview in Section~\ref{sec:experiment} shared a desire to utilize generative models not only as the core model (P5, 6, 8, 12), but also as a way to augment and simulate data (P7, 10, 11), showing the rapid pace of adoption of this new technology. 
In addition, governments around the globe including the US~\cite{white2023executive} and the EU~\cite{european2023ai} have been moving diligently towards introducing regulations and directives on AI, especially in recent years.
While we believe that the overall structure and contents of planning would remain less affected by the development of new models, we admit that our understanding and the workbook may need adaptation over time with the shifting AI landscape that will certainly continue.
For example, we hypothesize that information needs on the data may decrease as LLMs have enabled few-shot and zero-shot learning approaches~\cite{kojima2022large}.
We leave monitoring the effects of the endlessly shifting AI landscape on planning as well as making the required adaptations to the workbook as future work.

%% file: sections/08-conclusion.tex
\section{Conclusion}

Client-AI expert collaborations in AI application development are commonplace, and the planning stage is a key stage of this collaboration in which the client collects information about their goals, expectations, and resources to the AI expert for them to sketch out a technical approach to the development.
We examine the planning stage in AI application development in detail through an interview surveying 10 successfully concluded AI application development projects.
Through the interview study, we identify the steps of the planning stage along its three phases and deduce how a documentation form outlining the information needs of the AI expert from the client can facilitate the collaboration.
Next, we built a taxonomy of the AI experts' information needs and uncovered the characteristics of the AI development project that affect the information needs through an experiment on 12 AI experts.
Based on our findings, we introduce \wbname{}, a workbook designed to support effective collaboration between AI experts and clients.
Through two case studies, we find that the workbook not only can serve as an information collection and assessment guide for clients as they prepare for collaboration, but also as a discussion guide for AI experts that helps AI experts ensure the completeness of the information flow from the clients.
Although we verify the values of the workbook as it stands, the workbook will require continuous adaptation to the quickly changing AI landscape to provide maximal value to the client-AI expert collaboration in AI application planning.